\documentstyle[12pt]{article}
\title{
AB interface in superfluid $^3$He and Casimir effect.}
\author{ G.E. Volovik\\
Low Temperature Laboratory\\
Helsinki University of Technology\\
Otakaari 3A, 02150 Espoo, Finland\\
and\\
L.D. Landau Institute for Theoretical Physics, \\
Kosygin Str. 2, 117940 Moscow, Russia\\
}
\begin{document}
\maketitle
\begin{abstract}
{ The friction force on the moving interface between two different vacuum states
of superfluid 3He is considered at low temperature. Since the dominating
mechanism of the friction is the Andreev reflection of the massless
"relativistic" fermions, which live on the A-phase side of the interface, the
results are similar to that for the perfectly reflecting mirror moving in the
quantum vacuum.}
\end{abstract}

Submitted to JETP Letters, February 24, 1996.
\vfill \eject

\section{Introduction}

The AB interface is the boundary between two different superfluid vacua of
$^3$He. The dynamics of the interface is determined by the fermionic
quasiparticles (Bogoliubov excitations). In the A-phase vacuum the fermions are
chiral and massless, while in the B-phase vacuum they are massive. At the
temperature $T$ well below the temperature $T_c$ of the superfluid transtition
the thermal fermions are present only in the A-phase. Close to the gap nodes,
ie   at $\vec p \approx \pm p_F\hat l$, the  energy spectrum $E(\vec p)$ of the
gapless A-phase fermions  becomes "relativistic" \cite{Exotic}:
$$E^2(\vec p) = g^{ik}(p_i -eA_i) (p_k -eA_k)~~,\eqno(1.1)$$
where the vector potential is $\vec A=p_F\hat l$; $e=\pm$; and the metric
tensor is
$$g^{ik}= c_\perp^2 (\delta^{ik} - \hat l^i \hat l^k) +  c_\parallel^2 \hat l^i
\hat l^k ~~.\eqno(1.2)$$
Here  $\hat l$ is unit vector in the direction of the gap nodes in the
momentum space; $c_\perp= \Delta / p_F$ and $c_\parallel=v_F$ (with $c_\perp
\ll c_\parallel $) are  "speeds of light" propagating transverse to $\hat l$
and along $\hat l$ correspondingly;
$p_F$ is the Fermi momentum; $v_F=p_F/m_3$ is the Fermi velocity;
$m_3$ is the mass of $^3$He atom; $\Delta$ is the gap amplitude in
$^3$He-A.

In the presence of superflow with the superfluid velocity $\vec v_s$ the
following term is added to the energy $E(\vec p)$:
$$\vec p\cdot \vec v_s=(\vec p-e\vec A)\cdot \vec v_s  +e A_0~~,~~A_0=p_F\hat
l\cdot \vec v_s ~~.\eqno(1.3)$$
The second term corresponds to the scalar potential $A_0$ of the electromagnetic
field, while the first one leads to the nonzero element $g^{0i}=v_{s}^i$ of the
metric tensor and to the change of the elements $g^{ik}\rightarrow g^{ik}_{\rm
stationary}-v_{s}^iv_{s}^k$. As a result the Eq.(1.1) transforms to
$$ g^{\mu\nu}(p_\mu -eA_\mu) (p_\nu -eA_\nu)=0~~,\eqno(1.4)$$
with $g^{00}=-1$, $p_\mu=(\vec p,E)$, $A_\mu=(\vec A,A_0)$.

Since the B-phase excitations are massive, the A-phase excitations
cannot propagate through the AB interface. The scattering of the A-phase
fermions from the interface, which is known as Andreev reflection
\cite{Andreev}, is the dominating mechanism of the friction force experienced by
the moving AB interface. Due to the relativistic character of the A-phase
fermions the dynamics of the interface becomes very similar to the motion of the
perfectly reflecting mirror in relativistic theories, which was heavily
discussed in the relation to the Casimir effect (see eg \cite{Neto,Law}). So the
investigation of the interface dynamics at $T\ll T_c$ will give the possiblitity
of the modelling of the  effects of quantum vacuum.  On the
other hand, using the relativistic invariance one can easily calculate the
forces on moving interface from the A-phase heat bath in the limit of low $T$ or
from the A-phase vacuum at $T=0$. This can be done for any velocity of wall with
respect to the superfluid vacuum and to the heat bath. We discuss here the
velocities below the "speed of light" in $^3$He-A. The case of the velocity
exceeding $c_\perp$, which is rather typical in experimental situations
especially at low $T$ where the measured velocity of the interface is high, will
be discussed later.

\section{Force on moving wall at finite temperature:
Massless isotropic relativistic fermions.}

The motion of the AB interface in the so called ballistic regime for the
quasiparticles has been considered in \cite{YipLeggett,KopninAB,LeggettYip} (see
also \cite{Palmeri}). In this regime the force on the  interface  comes from
the mirror reflection at the interface (Andreev reflection) of the
ballistically moving thermally distributed Fermi particles.
Three velocities are of importance in this process: superfluid velocity of the
condensate $\vec v_s$, normal velocity of the heat bath $\vec v_n$ and the
velocity of the interface $\vec v_L$. The friction force is absent when the wall
is stationary in the heat bath frame, ie
$\vec v_L=\vec v_n$.

Let us first consider the nonrealistic model in which the speed of "light" is
isotropic, ie $c_\perp= c_\parallel=c$, and the vector potential
$\vec A$ is absent. In the next Section the results will be extrapolated to
the real AB interface. In the reference frame of the interface the system is
stationary thus the  energy of the quasiparticles in this frame
$$E'=E +(\vec v_s-\vec v_L)\cdot \vec p~~,~~E=cp~~,\eqno(2.1)$$
is conserved during the scattering. In thermal equilibrium
their distribution function is
$$f(\vec p)=  (1+e^{E'-(\vec v_n -\vec v_L)\cdot \vec p\over
T})^{-1}=(1+e^{E -(\vec v_n -\vec v_s)\cdot \vec p\over T})^{-1}
~~.\eqno(2.2)$$

Let us introduce the velocities with respect to the superflow, $v_L=v_L - v_s$,
and $v_n=v_n -v_s$.  Then the spectrum
in the frame of the wall  is $E'=cp -\vec p\cdot\vec v_L$ and the distribution
function in the frame of the wall is
$f(\vec p)=1/(1+e^{(cp -\vec p\cdot\vec v_n)/T})$.  In the ballistic
regime one calculates the momentum transfer from the heat bath to the wall due
to scattering at the wall
$$F=\sum_{\vec p} \Delta p_z {dE'\over dp_z} f(\vec p)~~.\eqno(2.3)$$
Here
$${dE'\over dp_z}=c\cos\theta -(v_L - v_s)~~\eqno(2.4)$$
is the group velocity of the particles in the wall frame;
$$\Delta p_z=2p {\cos\theta- (v_L - v_s)/c\over  1 -(v_L -
v_s)^2/c^2}~~.\eqno(2.5)$$
is the momentum transfer after reflection, where  $\theta$ is the angle between
the particle momentum $\vec p$ and the velocity of the wall $\vec v_L$. $\Delta
p_z$ is small compared to the cut-off parameter $p_F$, which correponds to the
Andreev reflection in condensed matter. The force per unit area is:
$${F(v_L-v_s,v_n-v_s)\over A}= - \hbar c {7 \pi^2 \over 60} {T^4\over (\hbar
c)^4}\alpha(u_L,u_n) ~~,~~u_L={v_L - v_s\over c}~~,~~u_n={v_n - v_s\over c}$$
$$\alpha(u_L,u_n)= {1\over 1-u_L^2}\int_{-1}^{u_L}  d\mu~  {(\mu
-u_L)^2 \over (1-\mu u_n)^4}
~~.\eqno(2.6)$$

Now we can consider several different cases.

\subsection{ $v_L\neq v_s=v_n$.}

In this most typical case the distribution function is the Fermi function
$f(E)=1/(1+e^{E/T})$, with $E=cp$. From Eq.(2.6) one has
$$\alpha(u_L,0)={1\over 1-u_L^2}\int_{-1}^{u_L}  d\mu~  (\mu -u_L)^2= {1\over 3}
+ u_L+ {4\over 3} {u_L^2\over 1-u_L} ~~.\eqno(2.7)$$
The force  disappears at $v_L-v_n\rightarrow -c$, because the particles cannot
reach the wall moving with the speed of light.  At $v_L=v_s=v_n$
the first term in the rhs of Eq.(2.7) gives  a conventional pressure $P$  on the
wall from the gas of particles, $F(v_L=v_s=v_n)=-AP$, where $A$ is the area of
the wall and
$$P=\hbar c {7 \pi^2 \over 180} {T^4\over (\hbar c)^4}~~\eqno(2.8)$$
The second term, which is linear
in $v_L-v_n$, is the friction force  on the moving wall if the wall moves
with respect to the heat bath:
$$F_{\rm friction}=-A\Gamma (v_L-v_n)~~,
~~\Gamma=\hbar c {7 \pi^2 \over 60} {T^4\over (\hbar c)^4}~~.\eqno(2.9)$$

\subsection{ $v_L= v_s\neq v_n$.}

The spectrum of the particles in the reference frame of the wall is
relativistic, $E'=cp$, while the distribution function is the Doppler shifted
Fermi function
$f(\vec p)=1/(1+e^{E'+(\vec v_s-\vec v_n)\cdot\vec p\over T})$.
From Eq.(2.6) one has
$$\alpha(0,u_n)= \int_{-1}^0  d\mu~  {\mu^2
\over (1-\mu u_n)^4}=  {1\over 3} (1-{v_L-v_n\over c})^{-3}
~~.\eqno(2.10)$$
For the small $v_L-v_n \ll c$ the results for the pressure and the friction
force are the same as in previous subsection. Difference occurs at higher
velocity: when $\vec v_s -\vec v_n$ approach $c$ the vacuum becomes unstable.

\subsection{  $v_L= v_n\neq v_s$.}

When the interface moves with the heat bath the force is an even function of
$v_L-v_s$:
$$\alpha(u_L=u_n)= {1\over 3}   (1-   {(v_L-v_s)^2\over c^2}  )^{-2}
~~.\eqno(2.11)$$
This means that the friction force is absent since the interface is in
equilibrium with the heat bath. The effect of the superflow  $\vec v_s-\vec v_L$
across the interface leads to the relativistic renormalization of the
temperature in the expression for the pressure:
$$T_{\rm effective}= {T \over \sqrt{g_{00}}}~~,~~g_{00} =   1-{v_s^2\over
c^2}  ~~,\eqno(2.12)$$
where the superfluid velocity is in the reference frame of the wall and heat
bath. This is in agreement with the Unruh analogy in which the superfluid
velocity plays the part of the gravitational potential \cite{UnruhSonic}.

\section{Force on moving AB interface at low $T$.}

Now let us apply the obtained results to the A-phase, which has an  anisotropic
velocity of light and also contains the vector potential $\vec A=p_F\hat l$. The
constant vector potential can be gauged away by shifting the momentum. If
$\vec p$ is counted from $e\vec A$ the situation is the same as in previous
Section with one exception: the Doppler shift leads also to the appearance of
the scalar potential: $A_0=\vec A\cdot
\vec v$. In the reference frame of the interface the  energy of
the quasiparticles becomes
$$E'=E +(\vec v_s-\vec v_L)\cdot \vec p +eA_0~~,~~E=\sqrt{g^{ik}p_ip_k
}~~,~~A_0=\vec A\cdot (\vec v_s-\vec v_L)~~.\eqno(3.1)$$

Since the scalar potential $A_0=const$, it does not influence the scattering of
the quasiparticles at the wall. The scalar potential can influence only the
thermal distribution function. But this does not happen in two cases: (i) when
$v_L\neq v_s=v_n$: in this case  the scalar potential arising from $\vec v_s$ is
compensated by the contribution from $\vec v_L$, and (ii) if $\hat l$ is
perpendicular to the flow the potential $A_0=0$. In both cases one has again the
thermal distribution function  $f(E)=1/(1+e^{E/T})$.

\subsection{ $v_L\neq v_s=v_n$.}

For the most symmetric solutions for the interface
structure the anisotropy vector $\hat l$ is either parallel or perpendicular to
the normal $\hat n$ to the wall (see Sections 3.14-15 in \cite{Exotic}). In both
cases the result for the force on the interface can be  obtained from the result
in previous subsection  by the rescaling of the momenta. Thus for
$v_L\neq v_s=v_n$ one has
$$F(v_L)_{\hat l \parallel \hat n}= - A\hbar {7 \pi^2 \over 60} {T^4\over
\hbar^4 v_F c_\perp^2} [ {1\over 3} + {v_L\over v_F} + {4\over 3}
 {v_L^2\over v_F^2}{v_F\over v_F- v_L }]~~.\eqno(3.2a)$$
$$F(v_L)_{\hat l \perp \hat n}= - A\hbar  {7 \pi^2 \over 60} {T^4\over
\hbar^4 v_F c_\perp^2} [ {1\over 3} + {v_L\over c_\perp} + {4\over 3}
{v_L^2\over c_\perp^2}{ c_\perp\over  c_\perp- v_L  }]~~.\eqno(3.2b)$$
Here $v_L$ is the velocity of the interface with respect to the heat bath.

In both cases the value of the pressure is the same, while the parameter
$\Gamma$
in the friction force is essentially different. The friction force for the case
$\hat l \parallel \hat n$ coincides with that obtained by Kopnin
in Ref.\cite{KopninAB}.  For $\hat l \perp \hat n$ the friction force is larger
by the factor $c_\parallel/ c_\perp\sim E_F/\Delta$.
For  these two directions the friction parameter and the pressure can be written
in the general form:
$$P=\hbar {7 \pi^2 \over 180} {T^4\over \hbar^4} (-g)^{1/2} ~~,
~~\Gamma=\hbar {7\pi^2 \over 60} {T^4\over \hbar^4} (-g)^{1/2} (g^{ik}n_i
n_k)^{-1/2}~~.\eqno(3.3)$$
Here $g_{\mu \nu}=(g_{ik}, g_{00}=-1)$ is the metric
tensor of the stationary A-phase with $g^{ik}$ from Eq.(1.2); $g=-1/(v_F^2
c_\perp^4)$ is the determinant of the metric tensor.

\subsection{ $v_L=v_n\neq v_s~$ , $~\hat l \perp \hat n$.}

Since the interface is stationary in the heat bath frame the friction force is
absent. Taking into account that for $~\hat l \perp \hat n$ the scalar
potential $A_0=0$ one obtains that the scattering of the fermions from the
interface  only to renormalizes the pressure:
$$P=\hbar {7 \pi^2 \over 180} ~{T^4\over  g^2_{00}\hbar^4}
(-g)^{1/2}={7 \pi^2 \over 180 \hbar^3}~ {T^4\over  v_F
c_\perp^2}(1-{v_s^2\over  c_\perp^2 })^{-2}~~
 ,\eqno(3.4)$$
where $v_s$ is the superfluid velocity in the heat bath frame.

\section{Casimir force on vibrating interface,  $T=0$.}

Let us now consider the case of the  oscillating  interface at $T=0$. For the
reflecting mirror in the form of the flat infinite plane oscillating in the
electromagnetic vacuum the result is as follows \cite{Neto}
$$\Gamma=\hbar {1 \over 60 \pi^2} {\omega^4 \over c^4 }  ~~,\eqno(4.1)$$
where $\omega$ is the frequency of oscillations.
For the Fermi vacuum the result is similar, vibrations of the interface lead to
the production of pairs  of fermions (see Refs.\cite{YipLeggett,LeggettYip}).
The friction force can be estimated  by extrapolation of the
 results in  Eqs.(3.2) for $T\neq 0$ if one substitutes $T \sim \hbar
\omega/\pi $ \cite{YipLeggett,LeggettYip}.  Let us find an exact expression for
the force using again the covariance of the fermionic spectrum of the A-phase.

The motion of the interface with alternating velocity leads to the time
dependence of the scalar potential $A_0$ in Eq.(1.3):
$$A_0(t)= p_F\hat l\cdot\vec v_s(t) ~~,\eqno(4.2)$$
where $\vec v_s(t)$ is the superfluid velocity in the reference frame of the
vibrating interface. This however has no effect since such time dependence can
be gauged away, ie compensated by the gauge transformation of the phase of the
wave function: $\phi(t) \rightarrow \phi(t) + e\int^t dt'A_0(t')$.

The effect of the alternating velocity $v_s$ comes from the time dependence
of the metric tensor
$$g^{i0}(t)=v_{s}^i(t)~~,~~g^{ik}(t)=g^{ik}_{\rm
stationary}-v_{s}^i(t)v_{s}^k(t)  ~~,\eqno(4.3)$$
with $g^{0i}(t)p_i=\vec p\cdot\vec v_s(t)$. If $p_z$ is a good quantum number,
then for each $p_z$ the time dependence can be compensated by the gauge
transformation, but due to the wall the momentum
$p_z$  is not conserved and this leads to mixing of states and finally to the
production of the pair of fermions. If the motion of the wall is periodic,
$\vec v_s(t)=\hat z v_{\omega}e^{-i\omega t}$, the term $\vec p\cdot\vec v_s(t)$
corresponds also to the action of the electromagnetic field with the
finite frequency $\omega$ but with zero wave vector. This field provides the
matrix element $M=  p_z v_{\omega}$ for the "photon" absorption. This
allows the annihilation of two particles, when they move to the wall. The energy
of the fermions is $E(\vec p_1)+\vec E(\vec p_2)=\omega$;  their transverse
momenta are opposite due to conservation of the momentum along the wall, $\vec
p_{1\perp}=-\vec p_{2\perp}$; the deficite of the momentum along the normal
$\hat n$ to the wall, $p_{1z}+p_{2z}\neq 0$, is absorbed by the wall. The
inverse process corresponds to the production of fermion pair from the vacuum
in the presence of the reflecting wall.

Let us consider first the case of isotropic fermions. The energy loss per unit
time due to the pair creation is
$$\Gamma_\omega v_{\omega}^2=$$
$$4  \omega\int {d^2p_\perp \over
(2\pi)^2}\int {dp_{z1}
\over  2\pi }
\int {dp_{z2} \over  2\pi }2\pi ({\vert M_1\vert^2\over \omega^2} +{\vert
M_2\vert^2\over \omega^2}) ~{\partial E_1\over \partial p_{z1} }~{\partial
E_2\over \partial p_{z2} }~
\delta(\omega - E(\vec p_1)-\vec E(\vec p_2))~~.\eqno(4.4)$$
Here $M_{1,2}=\vec p_{1,2}\cdot\vec v_{\omega}$; the factor $4$ takes into
account 2 spin species and two values of the "electric charge" $e=\pm 1$;
$\partial E  \over \partial p_{z}$ is the group velocity of the particle moving
towards the wall. Integration gives
$$\Gamma=\hbar {1 \over 30 \pi^2} {\omega^4 \over c^4 }  ~~.\eqno(4.5)$$
Extrapolating to the case of the anispotropic fermions in the A-phase one
obtains the friction parameter
$$\Gamma_{\hat l \parallel \hat n}= \hbar {1 \over 30 \pi^2} {\omega^4 \over
v_F^2 c_\perp^2 }  ~~,~~
 \Gamma_{\hat l \perp \hat n}= \hbar  {1 \over 30 \pi^2} {\omega^4 \over
 v_F  c_\perp^3 }  ~~.\eqno(4.6)$$
In the  case of the moving AB-interface this effect can be
observable since the velocities of "light" are small.

\section{Discussion.}

The relativistic description of the fermions in the A-phase of $^3$He allows
us to obtain easily many different results for the dynamics of the AB interface
in the low temperature limit. On the other hand there is one to one
correspondence between the motion of the interface and the Casimir effect for
the objects moving in quantum relativistic vacuum, which will alow to model
this effect in the experiments with the AB interface.

 The next steps are  (i) to extend calculations to the case  of arbitrary
angle between the normal of the interface and the orientation of the anisotropy
vector  $\hat l$; (ii) to find what happens when the velocity  of the
interface exceeds the smallest of the "speeds of light". This is
interesting especially at $T=0$, where some kind of Hawking radiation effect
should arise due to analogy between the superfluid velocity and the gravity
field.

I thank A.F. Andreev, A.J. Gill, T. Jacobson and N.B. Kopnin for illuminating
discussions. This work was supported through the ROTA co-operation plan
of the Finnish  Academy and the Russian Academy of Sciences and by the Russian
Foundation for Fundamental Sciences, Grants No. 93-02-02687 and 94-02-03121.

\vfill\eject

\end{document}